# A Method Using Photon Collapse and Entanglement to Transmit Information

**Authors and Affiliations**

Ling Hu[1],    Qiang Ni[1]*,

[1]School of Computing and Communications, Lancaster University; Lancaster, LA1 4YW, UK.

*Corresponding author.  Email: q.ni@lancaster.ac.uk

**Abstract**

Measurements cause quantum wave functions to collapse[1,2]. In tackling this elusive issue, we embark on the exploration of entropy exhibited by single-qubit quantum systems. Our findings surprisingly challenge the conventional law of entropy never diminishing[3,4]. We then interpret the confusing retrocausality phenomenon in Wheeler's delayed-choice experiments[5,6,7]. The entropy reduction and the quantum retrocausality can be combined to investigate how measurements lead to collapse - A close link is shown between quantum wave function collapse and the ubiquity of photons in the environments. Next, by studying the overlooked phenomena of quantum wave function collapse, we find that quantum eigenstate sets may be artificially controlled rather than randomly selected. Our study uncovers an often overlooked aspect of quantum wave function collapse - a potential avenue for deliberate manipulation of quantum eigenstate sets, deviating from the conventional notion of random selection. Leveraging this novel insight, we propose an innovative method for direct information transmission utilizing photon wave function collapse and entanglement. Given the lack of efficient approaches for employing quantum mechanisms in information transfer, our research aims to push the boundaries of quantum mechanics and contribute to advancing this field.

# Main

Quantum mechanics fundamentally changed our understanding of the micro-world because of its characteristics that go against normal cognition. Whether the microscopic particles behave as particles or waves depends on whether they are being measured - they show particle nature when measured; otherwise, they exhibit wave property[1].

A qubit is an abstract mathematical concept in quantum mechanics. It is the basic unit of information storage and processing. The qubit can appear in two different states simultaneously. It is called superposition, and these individual states are called eigenstates.

Let us consider a single-qubit quantum system built up with an electron. This electron can spin in an upward state and a downward state simultaneously. If we respectively record the upward and downward eigenstates as |0> and |1>, the superposition of the electron can be expressed as[2]:

$|\Psi>= \alpha |0> + \beta |1>$.

A measurement will cause the system's wave function to collapse from the superposition state to one of the eigenstates (|0> or |1>). The probability of obtaining the eigenstate |0> is $|\alpha|^2$, and that of getting the other eigenstate is $|\beta|^2$. We get $|\alpha|^2 + |\beta|^2 = 1$.

Let us start with the concept of entropy[3]. It is commonly associated with disorder, randomness, or uncertainty. It can be understood as the degree of the chaos of a system, which refers to the part of the energy that cannot be effectively invoked. The entropy $H$ is expressed as[4]:

$$H(X) = - \sum_{i=1}^{n} P(x_i) \log_2 P(x_i) ,$$

where $P(x_i)$ is the probability that the $x_i$ event will occur. Therefore, the entropy of the single-qubit quantum system is:

$$H = -|\alpha|^2 \log_2 |\alpha|^2 - |\beta|^2 \log_2 |\beta|^2 .$$

When this quantum system is at the state of superposition, which means $0 < |\alpha|^2 < 1$ and $0 < |\beta|^2 < 1$, then we get the entropy $H > 0$. On the other hand, after the wave function collapses, the uncertainty disappears, which means $|\alpha|^2 = 1$ ($|\beta|^2 = 0$) or $|\beta|^2 = 1$ ($|\alpha|^2 = 0$) cases. Either situation leads to the result of entropy $H = 0$.

The above result tells us that, after the collapse, the entropy of this system decreases from a positive value to 0. Some negative entropy (positive energy) must have been imported into this quantum system.

We know the measurement process cannot avoid involving some photons from surrounding environments. Part of them should have contacted the observed target electron. Therefore, these photons link this electron with the measurement process. Since the electron has no connection with other substances relating to measurement, we reasonably speculate that these photons from surrounding environments are the "agent photons" that take the function of transferring their negative entropy into this quantum system.

These "agent photons" may transfer their negative entropy to this electron in advance and then participate in the subsequent measurements. However, a question is created from the timeline with this interpretation - how do the "agent photons" know future events in advance and perform negative entropy energy transfer actions? To put it another way, is it possible that such a crash happens before the measurement?

To answer this question, let us look at Wheeler's delayed-choice experiment. It refers to a series of thought experiments proposed by John Archibald Wheeler (more than 40 years ago). Several experiments have been implemented to realize Wheeler's idea[5,6].

These experiments have fascinated many people. The results show that the observed particles play hide-and-seek with humans. They act as particles when the detectors can determine the "which-way" information; otherwise, they act as waves. If collapses are needed, they happen when choosing the paths, meaning that the observed particles collapse before they can be measured. These experiments overturn the law of causality in our brains. The effects precede the causes.

The Wheeler experiments have had many improved versions. A famous one among them was the delayed-choice quantum eraser experiment[7]. It had a complex structure and introduced entangled photon pairs into it. As people might expect, this ingenious experiment reconfirmed quantum retrocausality. That is, the measurements made on particles now can affect their properties in the past.

Quantum retrocausality has stunned countless people. So in the following, we need to interpret this phenomenon before utilizing it.

Special Relativity is a famous theory proposed by Albert Einstein in 1905[8]. The core of it is the Lorentz transformations[9]. Using these transformations, the time dilation effect is deduced. It means when an object moves at a very high velocity, its time passing slows down for an observer. The faster the movement, the slower the time passes.

Especially when the velocity reaches the speed of light $c$, time stops.

It means from the view of photons, time does not exist, and the world is static. Events do not happen according to the time sequence. They happen simultaneously for photons. In other words, since the time dilation effect reaches its extreme value, causality does not exist for photons.

Let us go back to the Special Relativity theory. We know it has upended many people's worldviews. The most puzzling question about it is, although Einstein took the constant speed of light (in a vacuum) as one of the two postulates of this theory, he did not explain why. Although many experiments have proven its correctness, it is still incomprehensible.

Considering that the constant speed of light means it is an independent parameter of our space coordinates. It has no relationship with our three-dimensional space's position, frequency, velocity, etc. Hence, the speed of light exhibits orthogonality to all these parameters. We believe that a reasonable interpretation of this orthogonality is that light operates in another dimension rather than the three-dimensional space. All we can see is only the projection of light.

Since the time order does not matter in front of photons, we can use the retrocausality characteristic to interpret the quantum wave function collapse process.

As we know, there are photons everywhere, and most of them are invisible. There are always some photons that both contact the observed target particles and participate in the measurement processes. These "agent photons" possess all the information on the timelines simultaneously. And they can act on the target particles when needed.

Regardless of the form of an observed target particle, such as a photon, an electron, or whatever, it needs negative entropy input to collapse its wave function. We know negative entropy is a type of positive energy, and energy cannot appear out of thin air. In addition, most observed target particles are not photons. They cannot reach the speed of light and thus cannot obtain global information. Therefore, they have neither the ability nor the will to collapse. They only passively accept negative entropy input to complete the collapse processes, and these "agent photons" are the party with the initiative. In case measurements exist (in the future), these "agent photons" transmit their negative entropy to the target particles at the proper places, causing the target particles' wave functions to collapse. The subsequent events are that these "agent photons" are annihilated within the measurements, such as being converted into electrical signals, erasing the traces they once had in this world.

The "agent photons" can control whether and where the target particles collapse. Where exactly does a collapse happen? It should be where the measurement starts to have anything to do with it. Take Wheeler's experiment as an example. When the observed particles start to choose "which way" to fly into, their wave functions collapse. That is, collapses may happen before the measurements.

The quantum wave function collapse seems far away from us, but it is hidden within our reach. In the following, we will introduce an optical experiment that has been repeated often over the past two hundred years. It includes the phenomena of quantum wave function collapse that goes unnoticed, which shows that the quantum eigenstate sets may be artificially controlled. The quantum wave function collapse is not as random as it has always been believed.

This experiment is usually used to interpret Malus law[10]. It exhibits some characteristics that are difficult to explain at the macro level. Occasionally, people ask why the results are so counter-intuitive, and the answers are always that they conform to the Malus law formulas, no problem. While this experiment can be explained with mathematical formulas, formulas are just descriptions. We still need the missing physical interpretations.

These miraculous properties should be caused by microcosm. We studied them and believe this experiment represents the transitions between various sets of quantum eigenstates. Moreover, these quantum transitions are manageable.

In Figure 1, the blue discs represent polarizers, and the red arrows represent the light intensity and directions. In Fig. 1(a), a natural light beam passes through two polarizers whose polarization directions are $\pi/2$ different. After the light beam goes through the first polarizer ($P_{11}$), its intensity reduces to half. The light beam disappears after it touches the second one ($P_{12}$).

**Fig. 1**: **Light polarization experiment.**

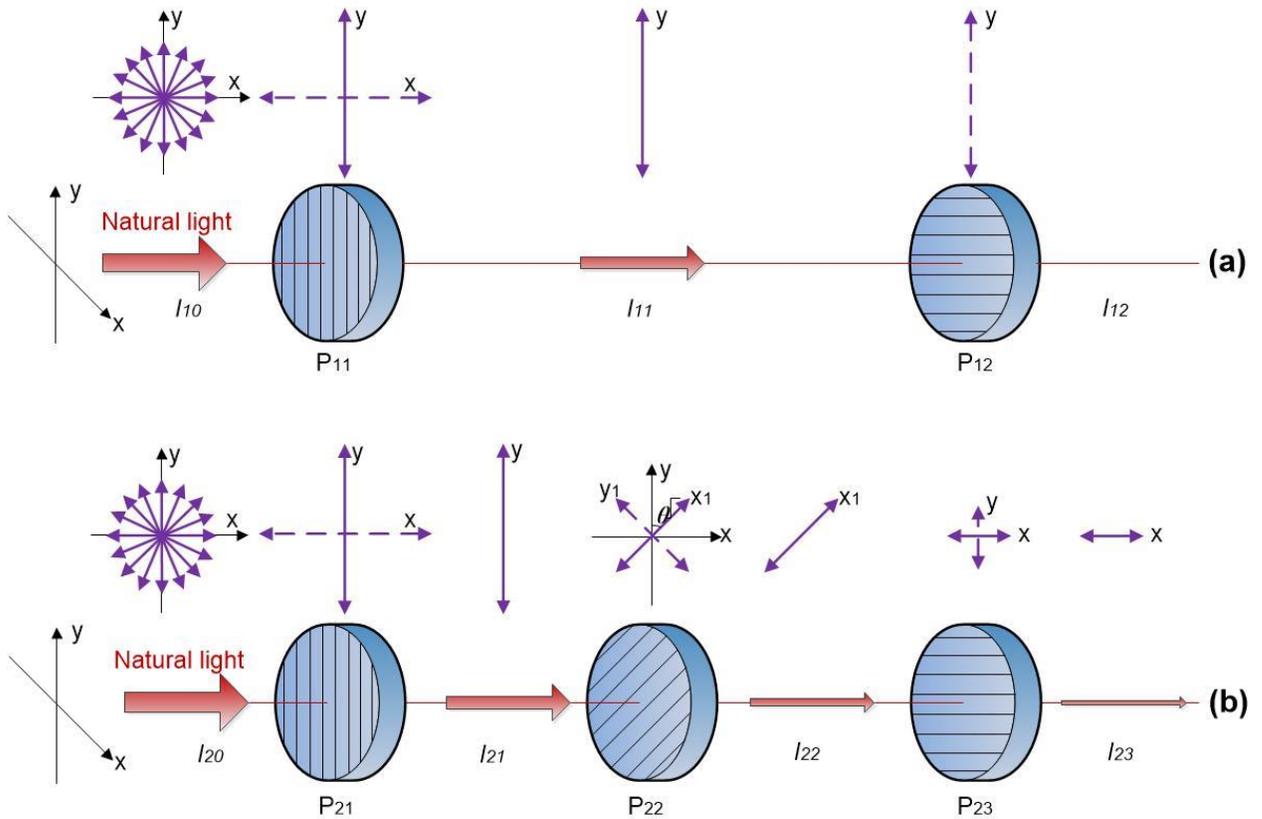

The blue discs represent polarizers, and the red arrows represent light intensity and direction. The purple vectors represent the polarization directions of the light. In experiment (a), a natural light beam passes through two polarizers whose polarization directions are $\pi/2$ different. After the light beam goes through the first polarizer ($P_{11}$), its intensity reduces to half. The light beam disappears after it touches the second one ($P_{12}$). After adding another polarizer between $P_{11}$ and $P_{12}$ to compose experiment (b), the three polarizers are renamed $P_{21}$, $P_{22}$, and $P_{23}$. The original opaque optical path surprisingly passes some light.

Then people add another polarizer between $P_{11}$ and $P_{12}$ to compose the experiment shown in Fig. 1(b). The original opaque optical path passes some light. Considering that the added polarizer is only a passive device, it should have brought more obstacles to this optical path and consumed more light energy. Why does some light pass?

It turns out that the light beams retain their color in the experiment. It means every photon keeps its frequency, wavelength, and energy unchanged. Therefore, changes in light intensity reflect the changes in photon number. Now let us first study the process of natural light passing through the first polarizer $P_{11}$ in Fig. 1(a). Experiment results demonstrate half of the light intensity left, and something in the micro-world emerges from this detail.

In this Figure, we use purple vectors to represent the polarization directions of the light beams. As we know, natural light shows isotropy in polarization directions. If the light polarization vibrates in only one direction, it is called linearly polarized light.

Since natural light behaves unpolarized, there are two possibilities for the photons of natural light. The first one is that each photon has its polarization direction, but all photons are equally distributed. The second one is that each photon stays in the superposition of all possibilities of polarization directions. Either case leads to the same result: a natural light beam shows unpolarized in the macro-world.

In the first case, if nothing changes in the microworld, the photons with exactly the y-axis (as shown in Figure 1) polarization direction would be only a tiny part of natural light. It is impossible to let as many as half of photons go through after being physically blocked by $P_{11}$.

Quantum mechanics tells us that when a quantum is not measured (disturbed), it should be in the superposition state. i.e., it is the second possibility - each photon that makes up natural light should be in the superposition of all the polarization directions. Following this idea, we speculate that the interaction between the natural light beam and a polarizer is a process for the photons that make up this beam to determine their eigenstate sets and collapse their wave functions.

Now let us focus on $P_{11}$. The photon wave functions collapse along the y-axis and x-axis. Since natural light is isotropic and unpolarized, the chances of collapsing in both directions are equal. Then the photons with the x-axis linear polarization directions are physically blocked by $P_{11}$. So the first result is that the passing through photons are linearly polarized in the y-axis direction. And the second result is that the light intensity $I_{11}$ is reduced to half of $I_{10}$.

The remaining photons continue to reach $P_{12}$. They are all blocked at last to make this optical path opaque.

After inserting the third polarizer between $P_{11}$ and $P_{12}$ to compose experiment (b), the three polarizers are renamed $P_{21}$, $P_{22}$, and $P_{23}$. $P_{22}$ has a polarization direction which makes the anger of $\theta$ with that of $P_{21}$. After a natural light beam passes $P_{21}$, linearly polarized light is gotten as before, and the light intensity is also $I_{21} = \frac{1}{2}I_{20}$.

Then the remaining light continues to interact with $P_{22}$. Since the polarization directions of the photons and $P_{22}$ are inconsistent, the photons should all be blocked. But the result is some light does pass through, and the passed light intensity $I_{22}$ is expressed as $I_{22} = I_{21}(\cos\theta)^2$ with Malus law. The linear polarization direction of the passing through light is rotated with anger $\theta$.

Why does the light pass with intensity decrease? We speculate that the photon eigenstate sets change directions with angle $\theta$, becoming $x_1$-$y_1$ axis directions as shown in the Figure. It means the photon wave functions collapse the second time as if the second measurement comes from $P_{22}$. The probability of collapsing to the $x_1$-axis direction can be calculated by the same cosine function consistent with the Malus law of macro-world. This time the photons with the $y_1$-axis polarization direction are physically blocked, resulting in some loss of light intensity.

When the photons continue to arrive at P$_{23}$, their wave functions collapse the third time on this optical path according to the third eigenstate set with the change of angle ($\frac{\pi}{2} - \theta$), becoming x-y axis polarization directions as in Figure 1. A portion of the photons are blocked as before; the final light intensity that passes through can be expressed as $I_{23} = I_{22}(\cos(\pi/2 - \theta))^2$. It is easy to prove that when $\theta = \pi/4$, the final transmitted light is the strongest. The maximum light intensity can reach the value as $I_{23} = \frac{1}{2}I_{22} = \frac{1}{4}I_{21} = \frac{1}{8}I_{20}$.

It is a fantastic result since this experiment goes against people's intuition. When Malus law was proposed more than two hundred years ago, quantum mechanics had not yet been established. That's why Malus did not realize what happened in the quantum world. However, after so many years, even though Malus law has been used as an explanation, it should be time to add microscopic physical interpretations to it.

It is known that the 2022 Nobel Prize in Physics was awarded jointly to Alain Aspect, John F Clauser, and Anton Zeilinger for their contributions to understanding quantum entanglement and advancing the field of quantum information[11,12,13].

However, there is an unsolved problem so far in the field of quantum information. Due to random collapse, there is no way to control the measurement results in the hands of the receivers after implementing quantum entanglement. If quantum wave function collapse can be intervened, utilizing quantum entanglement to transmit information directly may be possible.

As far as we know, entangled photons are commonly used sources for quantum entanglement[14]. If two photons are polarization entangled, their polarization directions are π/2 different: if one is a horizontally polarized photon, the other one is a vertically polarized photon[15].

In the above explanation about the optical experiment, a method of artificially controlling the photon eigenstate sets appears. Therefore, we devised the following experiment to use polarization-entangled photons to transmit information.

Let us use a coordinate axis system similar to that in Figure 1. First, generate a pair of polarization-entangled photons (A and B) and separate them far apart. Then let the photon A interact with a polarizer of the y-axis direction. In this way, the wave function of A will collapse to the x-axis or the y-axis. At the same time, the wave function of B will collapse to the y-axis or the x-axis accordingly. We define this case as the bit "1" being transmitted. Or if this polarizer rotates anger $\theta$ (for example, $\theta = \pi/4$), the wave function of A will collapse to the $x_1$-axis or the $y_1$-axis. At the same time, the wave function of B in the distance will collapse accordingly to the $y_1$-axis or the $x_1$-axis. We define this case as the bit "0" being transmitted. So the coordinate axes with different angles represent different transmission bits.

If we generate a series of polarization-entangled photon pairs to contact a polarizer that keeps rotating its angle, binary information can be transmitted continuously utilizing photon entanglement. At the receiving side, the polarizer directions of the collapsed photons need to be measured. In this way, we can receive the transmitted bits.

Since the polarization-entangled photon pair generation is inefficient and unstable at this stage, it is currently difficult to control the exact time of the photon pair generation. Progress depends on further developments in quantum technology.

In this article, we started with the entropy of a single quantum system and found that external energy is involved in the quantum wave function collapse process. Therefore, we combined it with quantum retrocausality to investigate the microscopic process of measurement-caused collapse. Next, we noticed an existing experiment showing the neglected phenomena of quantum wave function collapse, suggesting that photon wave function collapse could be intervened artificially. Therefore, we proposed an experiment for information transmission using photon wave function collapse and entanglement. The content flow chart for this article is shown in Fig. 2.

**Fig. 2: Content flow chart of this article.**

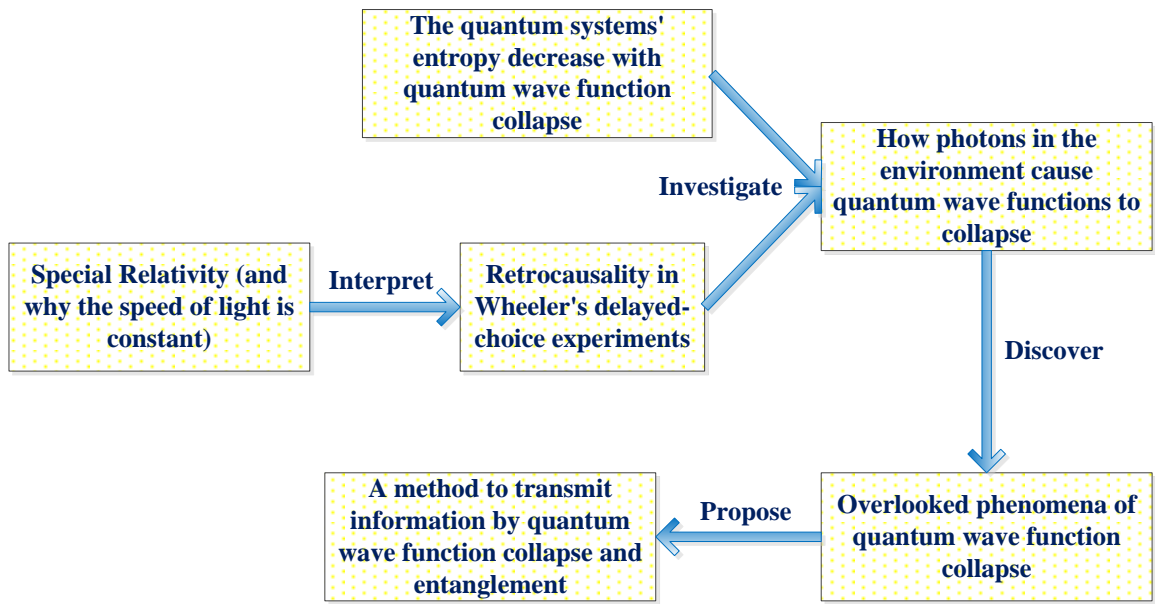

After finding entropy change in quantum systems and interpreting the retrocausality in Wheeler's delayed choice experiments, we investigated how photons in the environment cause quantum wave functions to collapse when measured. We then discovered the overlooked phenomena of quantum wave function collapse, which prompted us to propose a method to use photon wave function collapse and entanglement to transmit information.

Given the existing limitations in the effective use of quantum mechanics for information transfer, our research endeavors break new ground in quantum mechanics. By doing so, we aim to make contributions to the advancement of quantum science and its practical applications. We hope our work can provide valuable insights into this enigmatic field, leading to a clearer understanding of the intricate interplay between wave function collapse, quantum entanglement, and information transmission.